\documentclass{bioinfo}
\usepackage{inputenc}

\copyrightyear{XXXX}
\pubyear{XXXX}

\usepackage{natbib}
\usepackage{url}
\usepackage{alltt}
\usepackage{caption}
\usepackage{subcaption}
\usepackage{todonotes}
\usepackage{verbatimbox}

\newcommand\NGS{\textsc{Ngs}}

\newcommand\FASTA{\textsc{Fasta}}
\newcommand\FASTQ{\textsc{Fastq}}
\newcommand\ALN{\textsc{Aln}}
\newcommand\SAM{\textsc{Sam}}
\newcommand\ROC{\textsc{Roc}}
\newcommand\BAM{\textsc{Bam}}
\newcommand\HTML{\textsc{Html}}
\newcommand\RABEMA{\textsc{RABeMa}}

\newcommand\RNFTOOLS{\textsc{RnfTools}}
\newcommand\RNF{\textsc{Rnf}}
\newcommand\LAVENDER{\textsc{LAVEnder}}
\newcommand\MISHMASH{\textsc{MIShmash}}
\newcommand\SNAKEMAKE{\textsc{SnakeMake}}

\newcommand\CURESIM{\textsc{CuReSim}}
\newcommand\WGSIM{\textsc{WgSim}}
\newcommand\DWGSIM{\textsc{DwgSim}}
\newcommand\ART{\textsc{Art}}
\newcommand\PIRS{\textsc{Pirs}}
\newcommand\MASON{\textsc{Mason}}

\newcommand\XS{\textsc{Xs}}
\newcommand\FLOWSIM{\textsc{FlowSim}}
\newcommand\GEMSIM{\textsc{GemSim}}
\newcommand\PBSIM{\textsc{PbSim}}
\newcommand\SINC{\textsc{SINC}}
\newcommand\FASTQSIM{\textsc{FastqSim}}

\newcommand\CURESIMEVAL{\textsc{CuReSimEval}}
\newcommand\WGSIMEVAL{\textsc{WgSim\_Eval}}
\newcommand\DWGSIMEVAL{\textsc{DwgSim\_Eval}}

\begin{document}
\firstpage{1}

\title[RNF: a general framework to evaluate NGS read mappers]{RNF: a general framework to evaluate NGS read mappers}
\author[K. B{\v r}inda\ \textit{et~al.}]{
	Karel B{\v r}inda\,$^{1}$\footnote{to whom correspondence should be addressed},
	Valentina Boeva\,$^{2,3,4}$, and
	Gregory Kucherov\,$^{1}$
}
\address{
	$^{1}$LIGM/CNRS, Universit{\' e} Paris-Est, 77454 Marne-la-Vall{\' e}e, France;
	$^{2}$Inserm, U900, Bioinformatics, Biostatistics, Epidemiology and Computational Systems Biology of Cancer, 75248 Paris, France;
	$^{3}$Institut Curie, Centre de Recherche, 26 rue d'Ulm, 75248 Paris, France; and
	$^{4}$Mines ParisTech, 77300 Fontainebleau, France.
}

\history{Received on XXXXX; revised on XXXXX; accepted on XXXXX}
\editor{Associate Editor: XXXXXXX}
\maketitle

\begin{abstract}
	\section{Motivation:}
	Aligning reads to a reference sequence is a fundamental
	step in numerous bioinformatics pipelines. As 
	a consequence, the sensitivity
	and precision of the mapping tool, applied with certain
        parameters to certain data, 
	can 
critically affect the 
	accuracy of produced results (e.g.,
	in variant calling applications). Therefore, there has been an
	increasing demand of methods for comparing mappers
	and for measuring effects of their parameters.
	
	Read simulators combined with alignment evaluation tools
	provide the most straightforward way to evaluate and compare
	mappers. Simulation of reads is accompanied by information
	about their positions in the source genome. This
	information is then used to evaluate alignments
	produced by the mapper. Finally, reports containing 
	statistics of successful read alignments are created.

	In default of standards for encoding read origins, every
	evaluation tool has to be made explicitly compatible with the
	simulator used to generate reads.

	\section{Results:}
	In order to solve this obstacle, we have created
	a generic format \RNF\ (Read Naming Format) for assigning 
	read names with encoded information about original positions.

	Futhermore, we have developed an associated software package
	\RNFTOOLS\ containing two principal components.
	\MISHMASH\ applies one of popular read simulating 
	tools (among \DWGSIM,
	\ART, \MASON, \CURESIM\ etc.) and transforms
	the generated reads into \RNF\ format.
	\LAVENDER\ 
	evaluates then a given read mapper using simulated reads in
	\RNF\ format. A special attention is payed to
	mapping qualities that serve
	for parametrization of \ROC\ curves, and
	to evaluation of the effect of read sample contamination.
	
	\section{Availability and implementation:}
	
	\begin{tabular}{@{}ll}
	\RNF\ spec.: &
	\url{http://github.com/karel-brinda/rnf-spec} \\
	\RNFTOOLS: &
	\url{http://github.com/karel-brinda/rnftools} \\
	\end{tabular}

	\section{Contact:} \url{karel.brinda@univ-mlv.fr}
\end{abstract}

\section{Introduction}

The number of Next-Generation Sequencing ({\NGS}) read mappers
has been rapidly growing during the last years. Then,
there is an increasing demand of methods for evaluation and comparison of mappers in order to select the most appropriate one for a specific task.

The basic approach to compare mappers
is based on simulating {\NGS} reads, aligning them
to the reference genome and 
assessing read mapping accuracy using a tool evaluating
if each individual read has been aligned correctly.

There exist many read simulators
(\WGSIM\footnote{\href{https://github.com/lh3/wgsim}{http://github.com/lh3/wgsim}},
\DWGSIM\footnote{\href{https://github.com/nh13/DwgSim}{http://github.com/nh13/dwgsim}},
\CURESIM\ \citep{curesim},
\ART\ \citep{art},
\MASON\ \citep{mason},
\PIRS\ \citep{pirs}),
\XS\ \citep{xs},
\FLOWSIM\ \citep{flowsim},
\GEMSIM\ \citep{gemsim},
\PBSIM\ \citep{pbsim},
\SINC\ \citep{sinc},
\FASTQSIM\ \citep{fastqsim})
as well as many evaluation tools
(\WGSIMEVAL, \DWGSIMEVAL, \CURESIMEVAL, \RABEMA\ \citep{rabema}, etc.).
However, each read simulator 
encodes information about the origin of reads
in its own manner. This makes combining tools complicated and makes writing
ad-hoc conversion scripts inevitable.

Here we propose a standard for
naming simulated \NGS\ reads, called Read Naming 
Format (\RNF), that makes evaluation tools for read mappers
independent of the tool used for read simulation.

Furthermore, we introduce \RNFTOOLS, an easily configurable software, to obtain simulated
reads in \RNF\ format using a wide class of existing read
simulators, and also to evaluate \NGS\ mappers.

\subsection{Simulation of reads}

A typical read simulator introduces mutations into 
a given reference genome (provided usually as 
a \FASTA\ file)
and generates reads as genomic substrings with
randomly added sequencing errors.
Different statistical models can be employed to simulate 
sequencing errors and artefacts observed in experimental reads. The models usually take into account
CG-content, distributions of coverage, of sequencing errors in reads,
and of genomic mutations. Simulators
can often learn their parameters
from an experimental alignment file.

At the end, information about origin of every read
is encoded in some way
and the reads are saved into a \FASTQ\ file.

\subsection{Evaluation of mappers}

When simulated reads are mapped back
to the reference sequence
and possibly processed by an independent post-processing
tool (remapping around indels, etc.),
an evaluation tool inputs the final alignments of all reads,
extracts information about their origin and assesses if every single
read has been aligned to a correct location
(and possibly with correct edit operations). The whole procedure is finalized by creating
a summarizing report.

Various evaluation strategies can be employed
(see, e.g., introduction of~\citep{curesim}).
Final statistics usually strongly depends on
the definition of a correctly mapped read,
mapper's approach to deal with multi-mapped reads
and with mapping qualities.

\subsection{Existing read naming approaches}

Depending on the read simulator, information about read’s origin is either encoded in its name, or stored in a separate file, possibly augmented with information about the expected read alignment.
While \WGSIM\ encodes the first nucleotide of each
end of the read in the read name,
\DWGSIM\ and \CURESIM\ encode the leftmost nucleotide of each end.
\ART\ produces \SAM\ and \ALN\ alignment files,
\MASON\ creates \SAM\ files, and
\PIRS\ makes text files in own format. 

\begin{figure}[!tpb]
\scriptsize
\begin{subfigure}{1.0\linewidth}
\begin{verbatim}
Coor        12345678901234-5678901234567890123456789
                                                    
Source 1 - reference genome                        
chr 1       ATGTTAGATAA-GATAGCTGTGCTAGTAGGCAGTCAGCCC
chr 2       ttcttctggaa-gaccttctcctcctgcaaataaa     
                                                    
Source 2 - generator of random sequences                  
                                                    
READS:                                              
r001        ATG-TAGATA ->                           
r002/1         TTAGATAACGA ->                       
r002/2                                  <- TCAG-CGGG
r003/1                               tgcaaataa ->
r003/2              gaa-gacc-t ->                    
r004                       ATAGCT............TCAG ->
r005                                 GTAGG ->
             <- agacctt                           
                        <- TCGACACG   
r006       ATATCACATCATTAGACACTA
\end{verbatim}
\caption{Simulated reads}
\end{subfigure}

\begin{subfigure}{1.0\linewidth}
\scriptsize
\begin{tabular}{c|p{5.5cm}|c}
 \textbf{r. tuple} & \textbf{LRN} & \textbf{SRN} \\\hline
 r001
 	& \texttt{sim\_\_1\_\_(1,1,F,01,10)\_\_[single\_end]}
 	& \texttt{\#1}
 \\\hline
 r002
 	& \texttt{sim\_\_2\_\_(1,1,F,04,14),(1,1,R,31,39)\_\_\break[paired\_end]}
 	& \texttt{\#2}
 \\\hline
 r003
 	& \texttt{sim\_\_3\_\_(1,2,F,09,17),(1,2,F,25,33)\_\_\break[mate\_pair]}
 	& \texttt{\#3}
 \\\hline
 r004
 	& \texttt{sim\_\_4\_\_(1,1,F,15,36)\_\_[spliced],\break{}C:[6=12N4=]}
   	& \texttt{\#4}
 \\\hline
 r005
 	& \texttt{sim\_\_5\_\_(1,1,R,15,22),(1,1,F,25,29),\break(1,2,R,05,11)\_\_[chimeric]}
 	& \texttt{\#5}
 \\\hline
 r006
 	& \texttt{rnd\_\_6\_\_(2,0,N,00,00)\_\_[random]}
 	& \texttt{\#6}\\
\end{tabular}
\caption{Long and short read names}
\end{subfigure}
\caption{Example of simulated reads (in our definitions {\em read tuples}) and their corresponding
Long Read Names and Short Read Names, which can be used
as read names in the final \textsc{Fastq} file.
\label{fig:example}
}
\end{figure}

\section{Methods}

\begin{figure*}
	\centering

	\begin{subfigure}{.48\textwidth}
		\includegraphics[width=\textwidth]{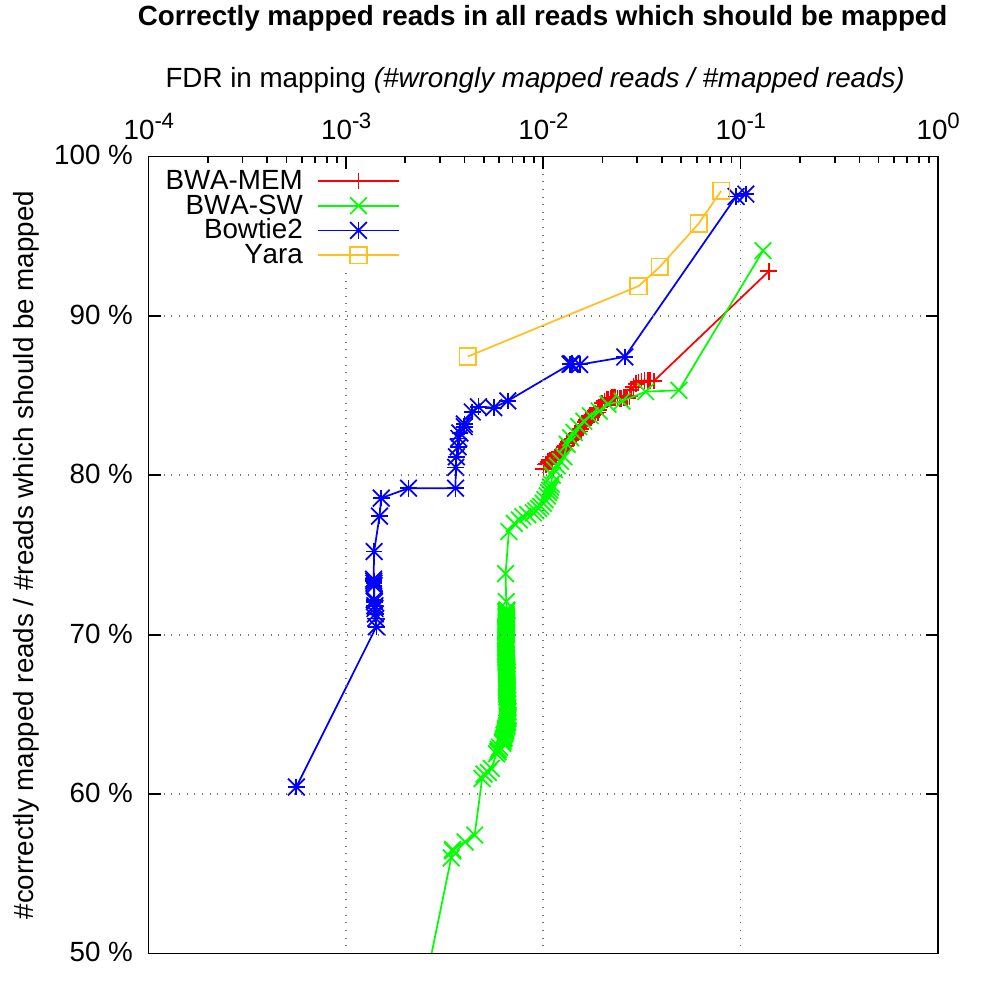}
	\end{subfigure}
	~
	\begin{subfigure}{.48\textwidth}
		\includegraphics[width=\textwidth]{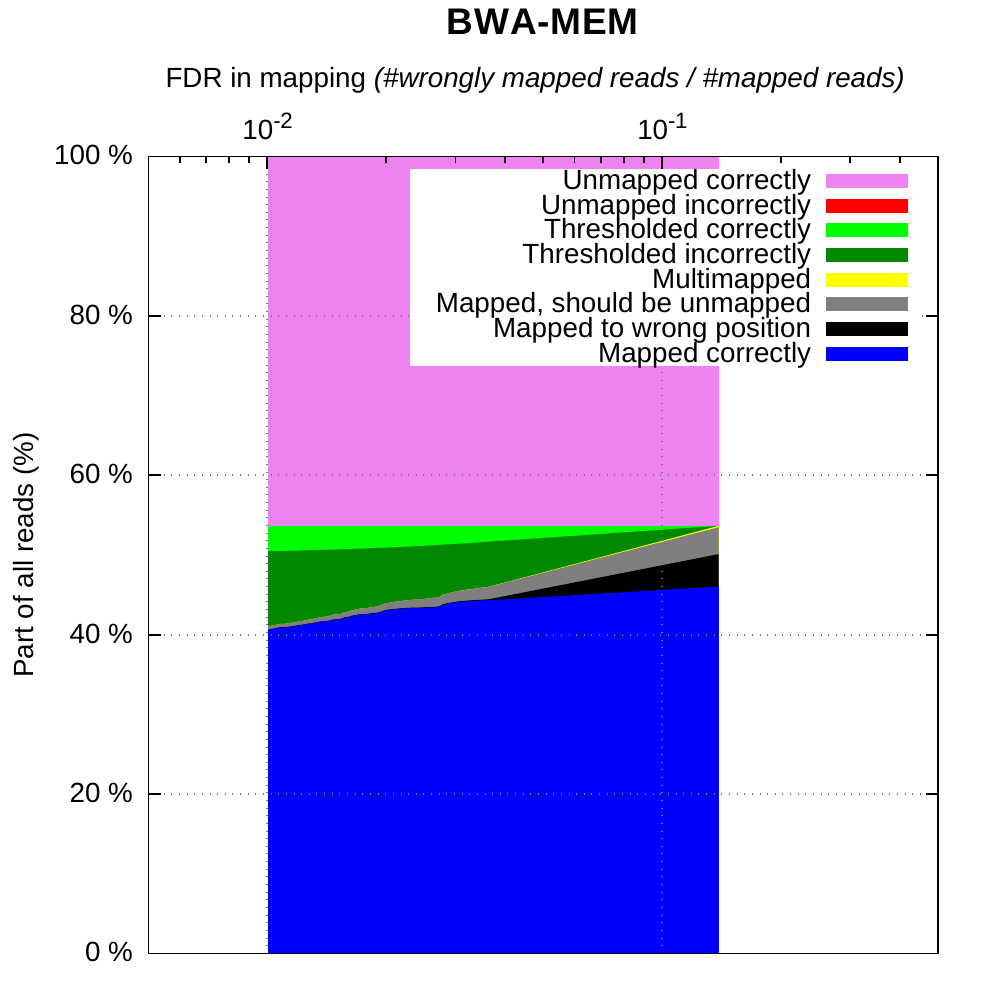}
	\end{subfigure}
	\caption{Example of two graphs produced by \LAVENDER\ as a part of comparison of mapper capabilities of contamination detection.
	$200.000$ single-end reads were simulated from human and mouse genomes ($100.000$ from HG38, $100.000$ from MM10) by \DWGSIM\ using \MISHMASH\ and mapped to HG38. All \LAVENDER\ graphs	have false discovery rate on $x$-axis and use mapping quality as the varying parameter for plotted curves.
	\label{fig:lavender}
	}
\end{figure*}

\begin{methods}
We have created \RNF, a standard for naming simulated reads.
It is designed to be robust,
easy to adopt by existing tools,
extendable,
and to provide human-readable read names.
We then developed an utility for generating \RNF-compliant
reads using existing
simulators, and an associated mapper evaluation tool.

\subsection{Read Naming Format (RNF)}

\subsubsection{Read tuples.}

{\em Read tuple} is a tuple of sequences (possibly overlapping) obtained from
a sequencing machine from a single fragment of DNA.
Elements of these tuples are called {\em reads}. For example, every ``paired-end read'' is a {\em read tuple} and both of its ``ends'' are individual {\em reads} in our notation.

To every {\em read tuple}, two strings are assigned:
a short read name (SRN) and
a long read name (LRN). SRN contains a hexadecimal unique ID of
the {\em tuple} prefixed by `\texttt{\#}`.
LRN consists of
four parts delimited by double-underscore:
i) a prefix (possibly containing expressive 
information for a user or a particular string
for sorting or randomization of order of tuples),
ii) a unique ID,
iii) information about origins of all segments (see below) that constitute {\em reads} of the {\em tuple},
iv) a suffix containing arbitrary comments or extensions (for holding additional information).

Preferred final read names are LRNs. If an LRN exceeds $255$ (maximum allowed read length in 
\SAM), 
SRNs are used instead and a SRN--LRN correspondence file must be created.

\subsubsection{Segments.}

{\em Segments} are substrings of a {\em read} which are spatially distinct in the reference and they correspond to individual lines in a \SAM\ file. Thus, each {\em read} has an
associated chain of {\em segments} and we associate
a {\em read tuple} with {\em segments} of all its {\em reads}.

Within our definitions,
a ``single-end read'' consists of a single {\em read} with a single {\em segment}
unless it comes from a region with genomic rearrangment.
A ``paired-end read'' or a ``mate-pair read'' consists
of two {\em reads}, each with one {\em segment} (under the same condition).
A ``strobe read'' consists of several {\em reads}.
Chimeric {\em reads} (i.e., reads corresponding to a genomic fusion, a long deletion, or a translocation) have at least two {\em segments}.

For each {\em segment}, the following information is encoded:
leftmost and rightmost $1$-based coordinates in its reference,
ID of its reference genome, ID of the chromosome and the direction ('F' or 'R').
The format is:\\
\texttt{(genome\_id,chromosome\_id,direction,L\_coor,R\_coor)}.

{\em Segments} in LRN are recommended to be sorted
with the following keys:
\texttt{source}, \texttt{chromosome},
\texttt{L\_coor}, \texttt{R\_coor}, \texttt{direction}. When some information
is not available (e.g., the rightmost coordinate), zero is used (`N' in case of direction).

\subsubsection{Extensions.}

The basic standard can be extended for specific purposes
by extensions (e.g., CIGAR string extension).
They are part of the suffix
and encode supplementary information.

Examples of \RNF\ names are shown in Figure~\ref{fig:example}.

\subsection{RNFtools}

We also developed~\RNFTOOLS, a software
package associated with~\RNF. It has two principal
components: \MISHMASH\ for read simulation and
\LAVENDER\ for evaluation of NGS read mappers.

\RNFTOOLS\ has been created using
\SNAKEMAKE\ \citep{snakemake},
a Python-based Make-like build system.
All employed external programs
are installed automatically when needed.

\subsubsection{MIShmash read simulator.}

\MISHMASH\ is a pipeline for simulating reads
using existing
simulators and combining obtained sets of reads together
(e.g., in order to simulate contamination or metagenomic
samples).
Its output files respect \RNF\ format, therefore, any \RNF-compatible evaluation tool can be used
for evaluation.

\subsubsection{LAVEnder evaluation tool for read mappers.}

\LAVENDER\ is a program for evaluating mappers.
For a given set of \BAM\ files,
it creates an interactive \HTML\ report with several graphs.

In practice, mapping qualities assigned
by different mappers to a given read are not equal
(although mappers tend to unify this). Moreover, even for
a single mapper, mapping qualities are very data-specific.
Therefore,
results of mappers after the same thresholding
on mapping quality are not comparable.
To cope with this, we designed \LAVENDER\ to use mapping qualities as
parameterization of curves in
`sensitivity-precision' graphs (like it has been done in~\cite{bwamem}).
Examples of output of \LAVENDER\ can be found in Figure~\ref{fig:lavender}.

\end{methods}

\section{Conclusion}

We designed \RNF\ format and propose it as 
a general standard
for naming simulated \NGS\ reads.

We developed \RNFTOOLS\ consisting of 
\MISHMASH,
a pipeline for read simulation,
and \LAVENDER, an evaluation tool for mappers, both following
the \RNF\ convention (thus inter-compatible).
Currently, \MISHMASH\ has a built-in interface with
the following existing read simulators:
\WGSIM, \DWGSIM, \ART, and \CURESIM.

We expect that authors of existing read simulators
will adopt \RNF{} naming convention as it is technically simple and
would allow them to extend the usability of their software.
We also expect authors of evaluation tools to use
\RNF\ to make their tools independent of the
used read simulator.

The main benefit for users is the ease of
switching between read simulators and
read evaluation tools,
without need of writing any special conversion scripts.
Altogether, it simplifies the process of evaluation of
\NGS\ mappers and accelerates the
debugging of tools for processing {\NGS} data.

\section*{Acknowledgements}

\paragraph{Funding\textcolon} This work was supported by
ABS4NGS grant and by Labex B\'ezout of the French government (program \emph{Investissement
  d'Avenir}). A partial
support has been provided by Labex B{\' e}zout grant. 

\paragraph{Conflict of Interest\textcolon} None declared.

\end{document}